\documentclass[aps,twocolumn,superscriptaddress,groupedaddress,10pt,nofootinbib]{revtex4} 

\usepackage{graphicx}  	
\usepackage{dcolumn}   	
\usepackage{bm}        	
\usepackage{amssymb}   	
\usepackage{amsmath}
\usepackage{epstopdf}
\usepackage[a4paper,margin=20mm]{geometry}
\usepackage{color}

\begin{document}

\title{Intrusion into granular media beyond the quasi-static regime}

\author{Leah K. Roth}
\affiliation{James Franck Institute and  Department of Physics, The University of Chicago, Chicago, Illinois 60637, USA}

\author{Endao Han}
 \thanks{Present address: Center for the Physics of Biological Function, Princeton University, Princeton, New Jersey 08544, USA}
\affiliation{James Franck Institute and  Department of Physics, The University of Chicago, Chicago, Illinois 60637, USA}

\author{Heinrich M. Jaeger}
 \email{E-mail: h-jaeger@uchicago.edu}
\affiliation{James Franck Institute and  Department of Physics, The University of Chicago, Chicago, Illinois 60637, USA}


\begin{abstract}
The drag force exerted on an object intruding into granular media can depend on the object’s velocity as well as the depth penetrated. We report on intrusion experiments at constant speed over four orders in magnitude together with systematic molecular dynamics simulations well beyond the quasi-static regime.  We find that velocity dependence crosses over to depth dependence at a characteristic time after initial impact. This crossover time scale, which depends on penetration speed, depth, gravity and intruder geometry, challenges current models that assume additive contributions to the drag.
\end{abstract}

\maketitle


Intrusion of a solid object into a static granular medium drives the material beyond yielding and triggers a transformation into a flowing state. 
Despite stress transmission along irregular, inherently heterogeneous pathways that trace out highly nonlinear interparticle contact forces \cite{behringer_particle_scale,Bi_2015,Hurley_2016}, the resulting resistance to intrusion can often be described in terms reminiscent of ordinary liquids \cite{Devaraj_2017_review,Kamrin_2016,Kondic_2012}. 
For sufficiently slow intrusion, the results are strikingly simple in that the drag force does not exhibit any velocity dependence at all and instead resembles an Archimedean law \cite{blumenfeld_nature, blumenfeld_soft_matter}. 
This quasi-static behavior forms the basis for recent models that predict the resistance to moving through or burrowing into granular material, situations important for animal and robot locomotion as well as for many engineering and geophysical applications \cite{Li_2009,Omidvar_2015, Kamrin_2016, goldman_robo, Goldman_2016_review}. 
When the intrusion speed is larger, a term is  added to account for inertial effects, and the drag force is typically represented by~\cite{Devaraj_2017_review, schiffer_vc, tsimring, durian_unified_force, umbanhowar_scaling, goldman_gran_imp_2010,Clark_2012, Clark_2013, durian_depth_dependent,Bester_2017_PRE} 
\begin{equation}
    F(z,v) = F_z\left(z\right) + F_v\left(v^2\right), 
    \label{eq:forceLaw}
\end{equation}
where $F_z$ is a linear function of  depth $z$ below the granular medium's free surface and $v$ is the velocity of the intruding object. 
The velocity component $F_v$ is assumed to have two distinct regimes, separated by a characteristic speed $v_c = \sqrt{2gd_\text{g}}$ associated with a grain of diameter $d_\text{g}$ settling under gravity $g$: 
for $v<v_\text{c}$, the system is quasi-static and $F_v$ is effectively zero; for $v>v_\text{c}$, $F_v \propto v^2$.

However, while these two separately depth- and velocity-dependent contributions to the drag force are used widely, their detailed physical justification has remained debated \cite{Devaraj_2017_review, blumenfeld_nature, durian_depth_dependent,durian_unified_force,gondret_exp_vel,melo_dynamics_shear}.  
Furthermore, whether the contributions are indeed additive has not been tested systematically. This is because of a prevalent focus on intrusion under constant force \cite{durian_unified_force, durian_depth_dependent, Katsuragi_2013,goldman_force_flow_prl, goldman_force_flow_pre, tsimring,Bester_2017_PRE}, which entangles the depth and velocity dependencies as the impacting object slows down and eventually comes to rest. 

To isolate the mechanisms, we here report on experiments and simulations performed with constant speed of intrusion. We focus on dry granular media with sufficiently large permeability, i.e., particle diameter, so that effects due  to interstitial air \cite{Devaraj_2017_review, Caballero_2007, Royer_2007} can be neglected. We find that Eq.~\ref{eq:forceLaw} does not properly capture the behavior of the drag force for $v>v_\text{c}$. 
Instead, beyond the quasi-static regime we observe a considerably more complex interplay: 
The drag force right after initial contact with the granular surface is velocity dependent but strongly peaked, and at later times (i.e. larger $z$) joins the pure depth dependence of the quasi-static limit, even if $v$ exceeds $v_\text{c}$ by as much as an order of magnitude.

\begin{figure}[b]
\includegraphics[width=\columnwidth]{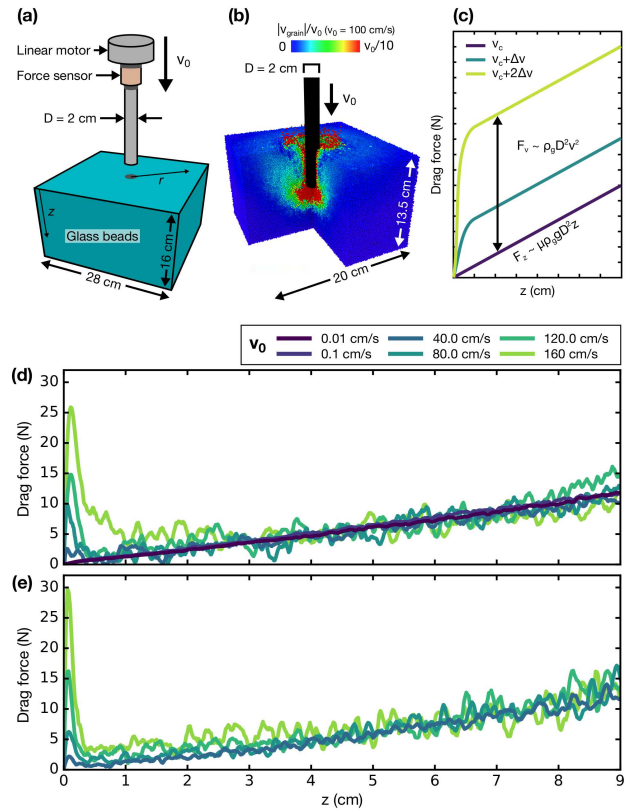}
\caption{\label{fig:setup_exp_sim} (a) Experimental setup. (b) Rendered image from simulation. Grains are colored by velocity ($v_0=100$ cm/s). (c) Expected results for the drag force on an intruder moving at a constant speed, based on existing models~\cite{durian_unified_force}. (d) Experimental and (e) simulation results for the drag force on a cylindrical intruder as a function of $z$ for different $v_0$.}
\end{figure}

In our experiments, a flat-bottomed, cylindrical rod with diameter $D=2$ cm was pushed into a granular medium, assembled via pouring and composed of glass spheres (Ceroglass, diameter $d_g \sim 2$ mm, density $\rho_g \sim 3$ g/cm\textsuperscript{3}), at a constant velocity $v_0$ that ranged from 0.1 mm/s to 2 m/s (Fig.~\ref{fig:setup_exp_sim}(a)).
The normal force applied by the particles on the rod during intrusion was measured by a force sensor aligned with the rod axis.
In order to visualize the particle dynamics during impact, molecular dynamics simulations were developed in LAMMPS~\cite{lammps} to reproduce, as closely as possible, the experimental system (Fig.~\ref{fig:setup_exp_sim}(b)).
For details about the methodology and the validation of the simulations, as well as for simulations performed using different rod diameters and gravitational accelerations and further details about the flow characteristics, see Ref.~\cite{Roth_arxiv}. 
In what follows, we discuss simulation results obtained with grains ($d_g = 2$~mm, $\rho_g = 3$~g/cm\textsuperscript{3}) that interact via a frictional Hertzian contact force, with normal stiffness $k_n = 10^9$~Pa and friction coefficient $\mu=0.3$, and that were prepared by pouring under gravity with a resulting packing density $\phi_0 \sim 0.61$~\cite{silbert_hertz}.

Based on Eq.~\ref{eq:forceLaw}, under conditions of constant  intruder speed, we would expect the $F-z$ curves to be parallel lines once $v_0$ exceeds the quasi-static regime, with a sharp transition only at the very beginning. This is sketched in Fig.~\ref{fig:setup_exp_sim}(c), where we write out the depth- and velocity-dependent contributions in terms of the system parameters, i.e. $F_z\left(z\right) \sim \mu \rho_g D^2 z$ and $F_v\left(v^2\right) \sim \rho_g D^2 v^2$.  

However, we find a velocity dependent peak in force at small $z$ that decays into a transient plateau regime before finally joining a common, linear $z$-dependence (Fig.~\ref{fig:setup_exp_sim}(d,e)).  Remarkably,  this linear depth dependence remains unchanged while  $v_0$ varies over four orders of magnitude, extending both below and above  $v_c$ (~$\sim20$ cm/s for this granular medium).
The simulations follow the experimental results in each of these respects, in addition to capturing the scale and frequency of the force fluctuations, which would likely decrease in magnitude as the ratio of $D/d_g$ increases.


\begin{figure}[h]
\includegraphics[width=\columnwidth]{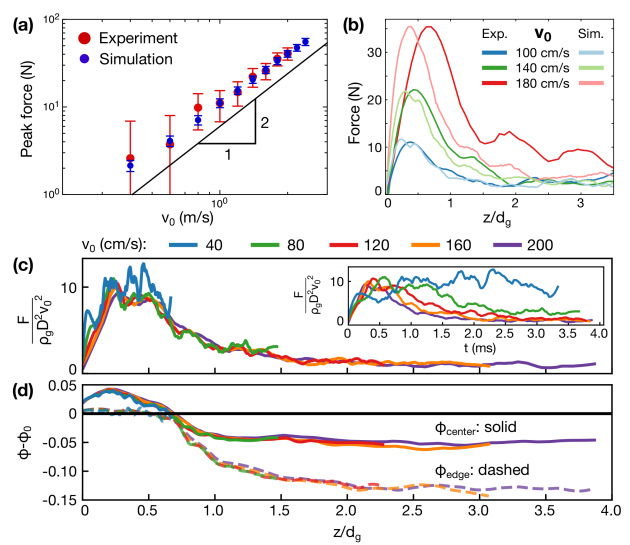}
\caption{\label{fig:peak_pf} (a) Peak force as function of $v_0$. (b) Peak morphology for three impact velocities. (c) Simulated drag force, scaled by $v_0^2$, as function of penetration depth. \textit{Inset:} Peak force as a function of time after impact. (d) Change in grain packing fraction directly below the center of the rod bottom (solid line) and below the edge of the rod bottom (dashed line).}
\end{figure}

\begin{figure}[h]
\includegraphics[width=\columnwidth]{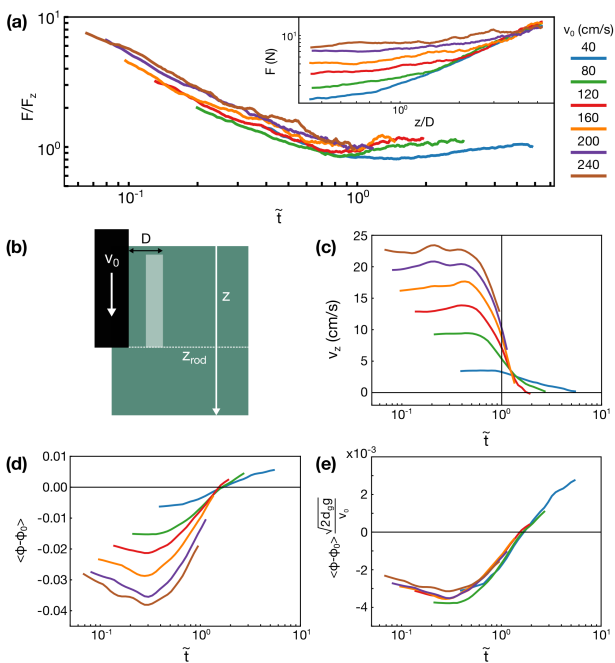}
\caption{\label{fig:pf_vz_avg_23}(a) Net drag force $F$ normalized by $F_z$, as function of rescaled time $\widetilde{t}$, excluding the initial peak. Inset: Net drag force as  function of intrusion depth. (b) Sketch indicating the region (highlighted) over which the quantities plotted in panels (c), (d) and (e) are averaged, an annulus extending vertically from $z=0.5$ to $z=z_{rod}$ with  inner radius $r=D$ and  outer radius $r=3D/2$, as measured from the rod axis. (c) Average vertical grain velocity. (d), (e) Average change in packing density. A return to the initial packing fraction $\phi_0$ coincides with the decrease in vertical velocity, indicating an initial fluidization that disappears at $\widetilde{t}=1$.}
\end{figure}

We first take a closer look at the initial force peak, which occurs as the intruding rod penetrates the top layer of grains. 
Direct comparison of the height (Fig.~\ref{fig:peak_pf}(a)) and profile (Fig.~\ref{fig:peak_pf}(b)) of the peak affirms the good agreement between simulation and experiment (the slight shift in simulated peak position is less  than half a grain diameter at the largest $v_0$).
The peak height is proportional to $v_0^2$ (Fig.~\ref{fig:peak_pf}(a)), a scaling  we  use  in Fig.~\ref{fig:peak_pf}(c) to collapse the force profiles also in the regime immediately following the peak.
Athani and Rognon recently also observed similar peaks in a 2d simulated system~\cite{athani}.
As the force peak occurs, the average local packing fraction directly below the rod bottom, $\phi_{\mathrm{center}}$, reaches a maximum that is independent of $v_0$ (Fig.~\ref{fig:peak_pf}(d)), while for further intrusion  both $\phi_{\mathrm{center}}$ and the local packing fraction at the edge of the rod, $\phi_{\mathrm{edge}}$, decrease below their original value.
This compaction followed by the onset of dilation comprises the initial nonlinear force regime, and indicates a yielding process that is similar to that described by Gravish \textit{et al.} in experiments where a plate was pushed horizontally through a granular medium~\cite{goldman_force_flow_pre}.

\begin{figure}[h]
\includegraphics[width=\columnwidth]{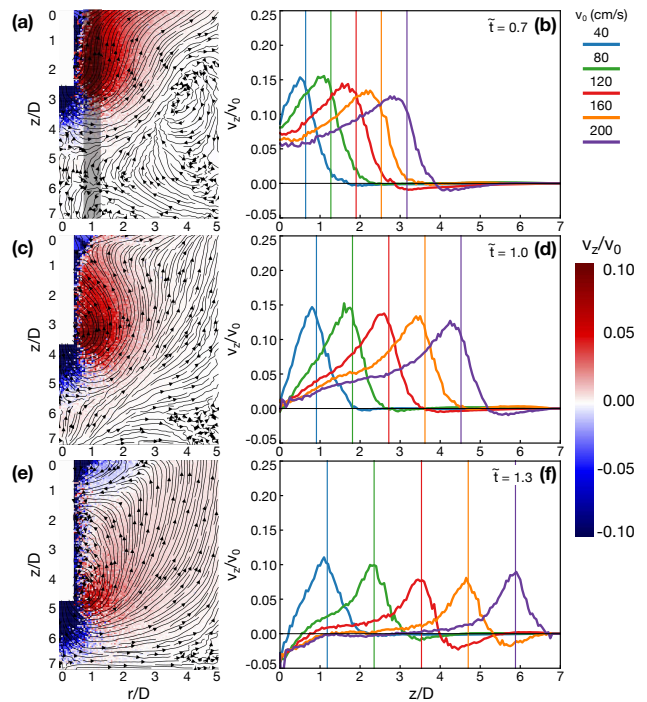}
\caption{\label{fig:vz_profile} Flow fields and vertical velocity component $v_z$ at times $\widetilde{t}=$  0.7, 1.0 and 1.3 (top to bottom rows). Left column, (a, c, e): Flow field in the $z-r$ plane for $v_0=160$ cm/s. Streamlines are shown in black, $v_z$ is indicated by color. The white rectangle in the upper left of each panel gives the rod position. Right column, (b, d, f): Vertical velocity profiles, averaged over the  strip shown shaded in (a), for five values of $v_0$. This strip extends from $r/D=0.75$ to $1.25$. Vertical lines delineate the rod bottom. Near the surface of the medium ($z \sim 0$) $v_z$ transitions from positive to negative once $\widetilde{t} > 1$.}
\end{figure}

\begin{figure}[h]
\includegraphics[width=\columnwidth]{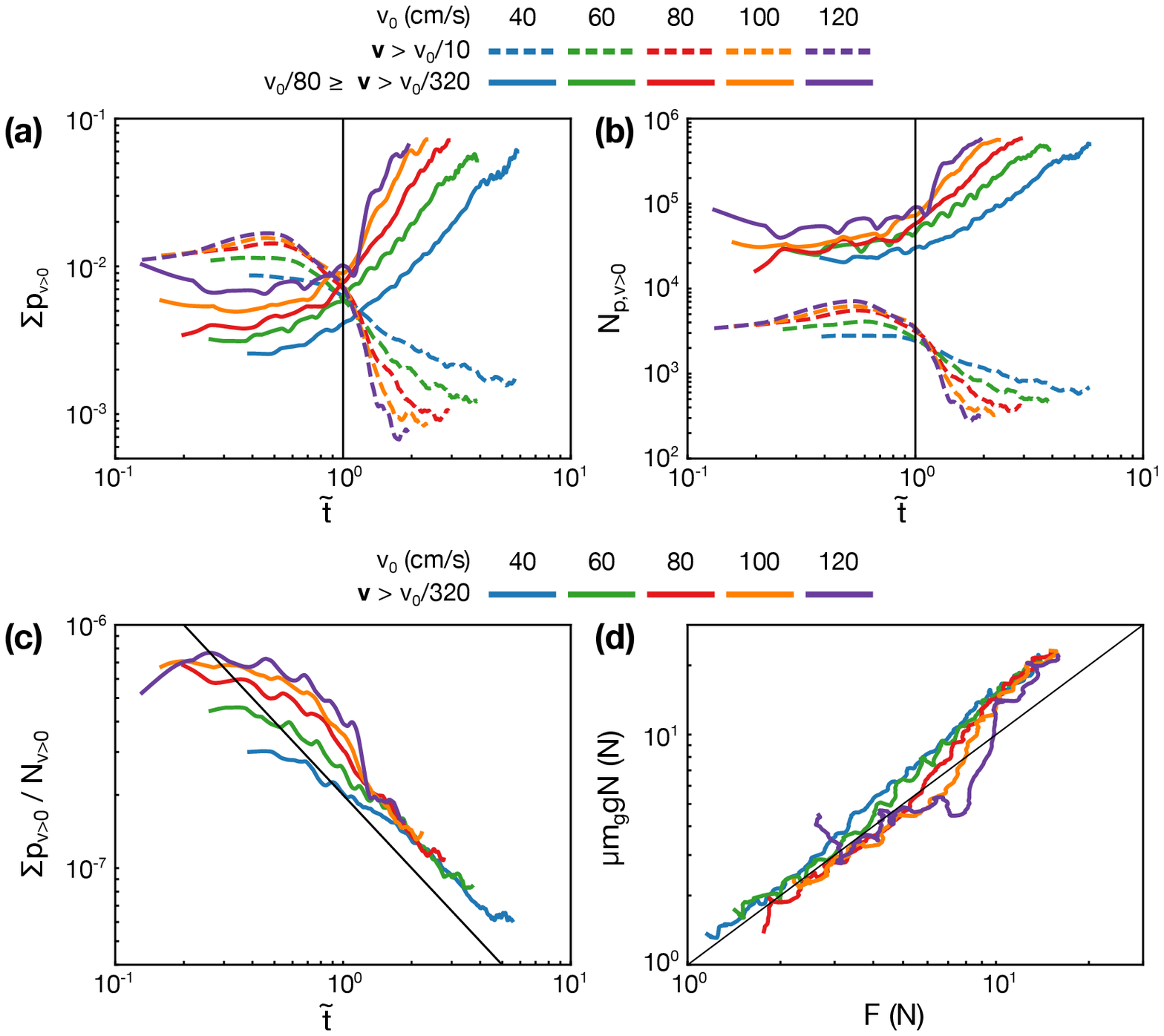}
\caption{\label{fig:mom_count} (a) Summed momentum of all grains located at $r>D/2$, binned by vertical velocity  magnitude. 
The momentum carried by the fastest grains ($v>v_0/10$, dashed line) decreases sharply at $\widetilde{t}=1$, while it increases for the slowest grains ($v_0/80 \ge v > v_0/320$, solid line). 
(b) The number of grains in the fastest bin also decreases at $\widetilde{t}=1$, while the number of slow grains grows. 
(c) The average momentum carried per grain decreases  throughout the intrusion and collapses  onto a  $v_0$-independent trend (the line indicates $1/\widetilde{t}$) for $\widetilde{t} > 1$. 
(d) Drag force on the rod plotted against the mass of all grains moving with an upward velocity of at least $v>v_0/320$ multiplied by the friction coefficient. A 1:1 line is plotted in black.}
\end{figure}

Following the initial peak is a short, transient region of effectively depth-independent drag force, before each trace joins a $z-$dependent behavior that has the form $F_z = \alpha \mu \rho_g g D^2 z$~\cite{durian_depth_dependent,durian_unified_force}.
Here the coefficient $\alpha$ takes on a value $35 \pm 5$, in line with  prior results for the quasi-static limit~\cite{durian_unified_force,durian_depth_dependent}. 
The force magnitude and extent in $z$ of this transient region depend on intrusion speed. 
This is most directly seen in the inset to Fig.~\ref{fig:pf_vz_avg_23}(a), where we excluded the initial peak and smoothed over the force fluctuations.
To isolate how the intrusion speed dependent drag crosses over to the depth dependent quasi-static behavior $F_z$, we show in the main panel of Fig.~\ref{fig:pf_vz_avg_23}(a) the overall drag force $F$ relative to $F_z$, again excluding the peak.
We find a collapse of the velocity dependent effects in the $F/F_z$ traces if we plot them  as
a function of dimensionless time
\begin{equation}
    \widetilde{t} = \frac{z}{v_0} \sqrt{\frac{g}{D}}. 
    \label{eq:time}
\end{equation}
For all velocities, $F/F_z$ is found to decrease linearly with $\widetilde{t}$ until, at $\widetilde{t}=1$, each force curve joins the quasi-static limit. Further data in support of the above functional form for $\widetilde{t}$ can be found in Ref.~\cite{Roth_arxiv}. 

We next examine the particle movements near the rod by means of a time dependent volume designed to capture general characteristics of the bulk flow, diagrammed in Fig.~\ref{fig:pf_vz_avg_23}(b).
The average vertical grain velocity $v_z$ in this region, though almost constant at early times, decreases sharply at $\widetilde{t}=1$ for all $v_0$ (Fig.~\ref{fig:pf_vz_avg_23}(c)).
Accordingly, the average packing fraction in this region decreases by up to four percentage points, before rapidly approaching, and even surpassing, $\phi_0$ at $\widetilde{t}>1$ (Fig.~\ref{fig:pf_vz_avg_23}(d)).
The packing fraction profiles for $\widetilde{t} < 1$ can be collapsed via the ratio of  impact speed $v_0$ to  quasistatic  velocity $v_c$ (Fig.~\ref{fig:pf_vz_avg_23}(e)).
This collapse implies that the fluidization of the medium, as evidenced by dilation as well as high vertical grain velocity, is both proportional to the speed of the intruder with respect to $v_c$, as well as limited in duration by the gravitational timescale $\widetilde{t}$.

To gain a more detailed perspective on the bulk flow over time, we plot in Fig.~\ref{fig:vz_profile} for three values of $\widetilde{t}$ the flow fields (a, c, e), as well as the averaged vertical velocity profiles $v_z(z)$ in a narrow region near the rod shaft (b, d, f).
At early scaled times ($\widetilde{t}=0.7$, Fig.~\ref{fig:vz_profile} top row), the flow is localized, upward along the side of the rod, and $v(z)$ at the surface of the granular medium is positive, signalling upwelling. 
At $\widetilde{t}=1.0$, we find signatures of a qualitative change in flow characteristics (Fig.~\ref{fig:vz_profile} middle row).
The stream lines have begun to bend back towards the rod, and there is no longer significant vertical motion at the surface.
This trend continues at later times (Fig.~\ref{fig:vz_profile} bottom row), with flow around the rod face becoming increasingly isolated within the bulk, while grains at the surface begin to fall back towards the fluidized volume.
Combined with Fig.~\ref{fig:pf_vz_avg_23}, this sequence illustrates a process of initial fluidization near the rod shaft, followed by eventual settling, beginning at $\widetilde{t}=1.0$, as the grains fall under gravity and collapse inwards.
This settling serves to isolate the flow around the rod's bottom face and produces an enhanced coupling between the intruder and the bulk, evidenced  in Fig.~\ref{fig:vz_profile}(e) by radially far extending flow lines paired with low vertical velocity, in contrast to the high velocity flows in Fig.~\ref{fig:vz_profile}(a).

Given these findings, we also expect a change in the behavior of the primary momentum carriers as a function of $\widetilde{t}$, especially during the transition away from high, localized mass flux at early times.
In Fig.~\ref{fig:mom_count}, we examine the relationship between momentum, vertical velocity, and $\widetilde{t}$ for the grains outside of the direct vertical path of the rod ($r > D/2$).
In panel (a), we divide these grains into two groups: those whose velocity is greater than $v_0/10$, and those whose velocity falls in the range $v_0/80 \ge v > v_0/320$.
At early times ($\widetilde{t}<1$) the majority of the momentum in the system is carried by the high velocity grains for all $v_0$, but at $\widetilde{t}=1$ a crossover occurs: the net momentum carried by high velocity grains decreases at the same time as that carried by very low velocity grains increases dramatically.

This change is also reflected in the absolute number of high and low velocity grains (Fig.~\ref{fig:mom_count}(b)).
Though the number of low velocity grains is larger than the number of fast grains at all times, as may be expected, at $\widetilde{t}=1$ the disparity begins to grow in the same way as the summed momentum.
Indeed, if we plot the average per-grain momentum for all grains with a minimum velocity $v > v_0/320$,
 we find that there is velocity dependence only for early times (Fig.~\ref{fig:mom_count}(c)), correlating with the fluidization of the bed observed in Fig.~\ref{fig:pf_vz_avg_23}(c-e).
At and after $\widetilde{t}=1$, the average momentum per grain collapses, independently of $v_0$, and 
appears to follow a $1/\widetilde{t}$ relationship from this point forward. This regime is associated with the isolation of the flow around the rod surface, seen in Fig.~\ref{fig:vz_profile}(e), especially in the growth of the flow lines as they penetrate into the
 previously uninvolved bulk grains far away from the rod surface.
 
It is also at $\widetilde{t}=1$ that the force on the rod transitions from exhibiting a velocity dependence to following the linear, quasi-static depth dependence  (Fig.~\ref{fig:pf_vz_avg_23}(a)).
Plotting the weight of all moving grains (taken here as those with $v > v_0/320$) multiplied by $\mu$, to approximate an effective frictional sliding force, against the force on the intruding rod shows how they are correlated (Fig.~\ref{fig:mom_count}(d)).
This agrees with aspects of the model proposed by Kang \textit{et al.}~\cite{blumenfeld_nature}, who find that the linear depth dependence is due to the engagement of an increasing number of particles with $z$. Our data extend this by showing that the force felt by the rod is proportional to the number of engaged particles even in the transient, velocity-dependent force region.

The findings can be summarized as follows.
At early scaled times ($\widetilde{t}<1$), the volume of grains near the rod surface has been fluidized in proportion to the ratio between $v_0$ and $v_c$ (Fig.~\ref{fig:pf_vz_avg_23}(d) \& (e)).
In this situation, the majority of the momentum generated via collisions with the rod's bottom surface is maintained by high velocity grains that travel through this fluidized region and thus lose little of their momentum (Fig.~\ref{fig:mom_count}(a)).
As the topmost grains decelerate under gravity and begin to settle, the mass flux of grains through the surface of the bed near the rod shaft decreases (Fig.~\ref{fig:vz_profile}(b, d, f)) and the stream lines fold backwards (a, c, e).
The flow around the rod surface is now isolated within the bulk, and the dilated and displaced grains near the rod tip must transfer their momentum more broadly, through many collisions with neighbors.
At this stage, the force felt by the rod joins with the linear quasi-static depth-dependent limit and loses its velocity dependent characteristics (Figs.~\ref{fig:setup_exp_sim} \& ~\ref{fig:pf_vz_avg_23}).
This behavior is not captured by Eq.1 and points to a need for revisiting the physical mechanisms for the additive drag model~\cite{tsimring,goldman_gran_imp_2010,umbanhowar_scaling,goldman_robo,durian_unified_force,blumenfeld_soft_matter}.
Much of this will entail a proper accounting of the relationship between $F_v$ and $F_z$, but our results indicate that $F_z$ is not linked to the presence of a stagnant zone underneath the intruder as was recently proposed~\cite{blumenfeld_nature}. Instead, we identify this as a region of pronounced dilation and shear that interacts with an extensive flow field reaching to the surface of the bed (Fig.~\ref{fig:pf_vz_avg_23})~\cite{goldman_force_flow_prl,goldman_force_flow_pre,tanaka_plate_drag,gondret_exp_vel}.

In essence, the whole dynamics is transitioning from an inertial to a quasi-static regime: whereas at early times, momentum is carried by localized, high velocity grains, at later times the energy injected into the system is spread out into the bulk and grains move quasi-statically (Fig.~\ref{fig:mom_count}(c)).
As a result, it appears that the velocity dependent characteristics of both the force felt by the rod and the flow of the displaced grains rely on access to the free surface of the bed as a way of maintaining the inertial qualities of the grains.
Once the flow field is isolated from the surface due to the gravitational settling of the initially fluidized region, the increasing number of grains that must be involved in the motion of the intruder 
qualitatively changes the nature of the flow.

We thank Kieran Murphy, Melody Lim and Abhinendra Singh for many helpful discussions, and Paul Umbanhowar for generously lending us glass spheres for this project.
This work was supported by the Center for Hierarchical Materials Design (CHiMaD), which is supported by the National Institute of Standards and Technology, US Department of Commerce, under financial assistance award 70NANB14H012, and by the Army Research Office under Grant Number W911NF-19-1-0245.






\bibliographystyle{unsrt}
\bibliography{const_speed_arXiv}

\end{document}